\documentclass[prl,twocolumn,aps,psfig,epsf]{revtex4}

\usepackage{graphicx}

\begin{document}

\title{Induced Ferromagnetism due to Superconductivity in
Superconductor-Ferromagnet Structures. }
\author{F.S. Bergeret$^{1,4}$, A.F. Volkov$^{1,2}$ and K.B.Efetov$^{1,3}$}
\address{$^{(1)}$Theoretische Physik III,\\
Ruhr-Universit\"{a}t Bochum, D-44780 Bochum, Germany\\
$^{(2)}$Institute of Radioengineering and Electronics of the
Russian Academy\\
of Sciences, 103907 Moscow, Russia\\
$^{(3)}$L.D. Landau Institute for Theoretical Physics, 117940
Moscow, Russia\\
$^{(4)}$ Laboratorio de F\'{i}sica de Sistemas Peque${\it
\tilde{n}}$os y Nanotecnolog\'{i}a, CSIC, Serrano144, E-28006
Madrid\\}

\begin{abstract}
We consider a superconductor-ferromagnet (S/F) structure and
assume that above the superconducting transition temperature
$T_{c}$ the magnetic moment exists only in F. {In a simple model
of the ferromagnet (the exchange field is of the ferromagnetic
type for all energies) }we show by an explicit calculation that
below $T_{c}$ the magnetic moment may penetrate the
superconductor. {In this model} its direction in S is opposite {to
the magnetization of free electrons} in the ferromagnet. The
magnetization spreads over a large distance which is of the order
of the superconducting coherence length $\xi _{S}$ and can much
exceed the ferromagnet film thickness. At the same time the
magnetic moment in the ferromagnet is reduced. This inverse
proximity effect may explain the reduction in magnetization
observed in recent experiments and may lead to a strong
interaction between the ferromagnetic layers in F/S/F structures.
\end{abstract}

\maketitle

\address{$^{(1)}$Theoretische Physik III,\\
Ruhr-Universit\"{a}t Bochum, D-44780 Bochum, Germany\\
$^{(2)}$Institute of Radioengineering and Electronics of the Russian Academy\\
of Sciences, 103907 Moscow, Russia\\
$^{(3)}$L.D. Landau Institute for Theoretical Physics, 117940 Moscow, Russia\\
}

Penetration of the superconducting condensate into a normal metal
in the superconductor (S)-normal metal (N) heterostructures is a
well established proximity effect. The latter is a long range
effect because the amplitude of the condensate decays in the
normal metal very slowly with a characteristic
length $\xi _{N}$ which in the dirty limit is equal to $\xi _{N}=\sqrt{%
D_{N}/2\pi T}$ ( $D_{N}$ is the diffusion coefficient in the
normal metal and $T$ is the temperature). At low temperatures this
length can be very large. At the same time, the order parameter
$\Delta $ in the superconductor near the S/N interface is
suppressed. The magnitude of the suppression depends on the
parameters characterizing the system such as the S/N interface
transparency, the thickness of the S and N layers, etc.\cite
{deGennes}. The proximity effect arises also in
superconductor-ferromagnet (S/F) structures. While the
superconducting condensate consists of paired electrons with
opposite spins, the exchange field $J$ in the ferromagnet tends to
align them and break the Cooper pairs. The penetration length $\xi
_{F}$ of the condensate into the ferromagnet is usually much smaller than $%
\xi _{N}$ and in the dirty limit is equal to $\xi _{F}=\sqrt{D_{F}/J}$ {\ ($%
D_{F}$ is the diffusion coefficient in the ferromagnet)}. Since
the exchange energy $J$ is much larger than $T$, we come to the
inequality $\xi _{F}\ll \xi _{N}$ (in the clean limit when $\tau
J>>1$, the penetration length $\xi _{F}$ is determined by the mean
free path $l=v_{F}\tau ,$ where $v_{F}$ is the Fermi
velocity\cite{Bulaev,Buzdin,BVE1}). A strong exchange field
suppresses also the superconducting order parameter $\Delta $ in
the superconductor.

The situation changes when the magnetization $\mathbf{M}$ in the
ferromagnet is not homogeneous. In this case a triplet component
of the condensate with a non-zero spin projection arises and
penetrates into the $F$- region over a long distance of the order
$\sqrt{D_{F}/2\pi T}$ {\ (see Refs. \cite {BVE2,Sweden}).} The
effect of the penetration of the superconducting condensate into
the ferromagnet and the suppression of the superconductivity (a
decrease of the critical temperature $T_{c}$ of the
superconducting transition) in $S/F$ structures, \textit{i.e.} the
proximity effect, has been the subject of many works during the
last decades (see for example \cite {radovic,demler,valls} and the
review \cite{Proshin} for more references).

So, the penetration of the superconductivity into the normal metal
or ferromagnet is by now a very well studied phenomenon. However,
one can ask the same question about the ferromagnetism: Can the
ferromagnetic order penetrate the normal metal or superconductor
over long distances? Surprisingly, this question has hardly been
addressed. Some indications of the effect can be found in
numerical works Refs. \cite{valls,fazio}. In those works only the
density of states for each spin direction as a function of the
energy was presented, however the magnetization was not
calculated. In addition the induced magnetization (
''magnetization leakage'') was calculated in Ref. \cite{KrKo}.
However the results obtained in the latter paper {generally
speaking} differ drastically from ours. They found a
''magnetization leakage'', that is the magnetic moment {of free
electrons} $M_{e}$ spreads into the S region over a distance of the order $%
\xi _{S}$ changing its sign at some distance from the S/F
interface. {\ }

{{We consider a simple model assuming a mean field approximation
for the ferromagnet and superconductor. The mean field order
parameter in S is
the energy gap }}$\Delta ,$ {and in F it is the exchange field }$J$%
{\ which is assumed to be of the ferromagnetic type and small
compared with the Fermi energy. }{In different limiting cases
where analytical formulae can be obtained} we find completely
different behavior:
For temperatures below $T_{c}$, the magnetization {of free electrons }%
in the F layer $M_{e}$ decreases and the induced magnetization in
the S region is \textit{negative} (that is, the magnetization
variation has the same negative sign in both regions). Our
analytical considerations show that no change of sign of the
induced magnetization takes place. This behavior is in agreement
with the reduction of magnetization observed in the experiments of
Refs.\cite{muhge,Garif} and can be explained by the simple
physical picture we present below.

\begin{figure}
\includegraphics[scale=0.4]{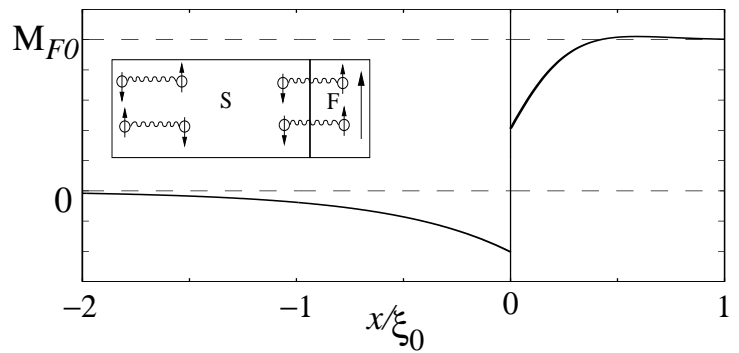}
\caption{Spatial dependence of the
magnetization in the whole system . Here $\gamma_F/\gamma_S =0.5$,
$\bar{\gamma}_F=\gamma_{F}/\xi_0=0.1$
($\xi_0=\sqrt{D_{S}/2T_{c}}$), $J/T_{c}=15$ and $d_{F}/\protect\xi
_{0}=1$. Inset: Schematic view of the inverse proximity effect in
a S/F system (for discussion see text).} \label{Fig.1}
\end{figure}

In the case of F/N systems the ferromagnetic ordering penetrates
over short distances since the exchange interaction is local. In
this paper we show that the situation may be different for $S/F$
structures and present arguments that the magnetic moment can
penetrate into the superconductor over long distances of the order
of the superconducting coherence length. This effect can be called
the inverse proximity effect. The reason why the
magnetic moment aligned in the direction opposite to the magnetization%
{\ }in the F film $\mathbf{M}_{e}$ penetrates the superconductor
can rather easily be understood qualitatively. This effect is due
to the fact
that the Cooper pairs have a large size of the order of $\xi _{S}\cong \sqrt{%
D_{S}/2\pi T_{c}}$. Suppose that the F layer is thin (see inset of
Fig. \ref {Fig.1}) and let us assume that the Cooper pairs are
rigid objects consisting of electrons with opposite spins, such
that the total magnetic moment of a pair is equal to zero. Of
course, the exchange field should not be very strong, otherwise
the pairs would break down. It is clear from this simple picture
that pairs located entirely in the superconductor cannot
contribute to the magnetic moment of the superconductor because
their magnetic moment is simply zero. Nevertheless, some pairs are
located in space in a more complicated manner: one of the
electrons of the pair is in the superconductor, while the other
moves in the ferromagnet. These are those pairs that create the
magnetic moment in the superconductor. The direction along the
magnetic moment $\mathbf{M}_{e}$ in the ferromagnet is preferable
for the electron located in the ferromagnet and this makes the
spin of the other electron of the pair be antiparallel to
$\mathbf{M}_{e}$. This means that all such pairs equally
contribute to the magnetic moment in the bulk of the
superconductor. As a result, a ferromagnetic order is created in
the superconductor and the direction of the magnetic moment in
this region is opposite to the direction of the magnetic moment $\mathbf{M}%
_{e}$ in the ferromagnet. Moreover, the induced magnetic moment
penetrates over the size of the Cooper pairs $\xi _{S}$. {\ From
this point of view it is difficult to understand the numerical
results of Ref. \cite{KrKo} where the induced magnetization in the
S region near the S/F interface has the same sign as in the
ferromagnet. The magnetic moment in the ferromagnet is decreased
because the density-of-states in F is reduced due to the proximity
effect. This occurs in a way similar to a suppression of the Pauli
paramagnetism in superconductors (the exchange field plays the
role of a strong magnetic field acting on spins). }{At the same
time the concentrations of free electrons with spin up and down in
F remain unchanged; when we are saying about the penetration of
Cooper pairs into the ferromagnet F, we mean that superconducting
correlations are established in F due to the proximity effect. }

Having presented the qualitative picture, we calculate now the
magnetization variation {due to free electrons (the conduction
band electrons)}
\begin{equation}
\delta M_{e}=\mu _{B}\delta N_{M}  \label{e1}
\end{equation}
below $T_{c}$ in both layers of the S/F system shown in the inset
of Fig.\ref {Fig.1}. Here $\mu _{B}$ is an effective Bohr
magneton. We assume that the magnetic moment $\mathbf{M}_{e}$ is
{\ parallel to the interface as it takes place in the experiment
\cite{muhge,Garif}} and is homogeneous in the $F$ layer. As we
have found previously, Ref.\cite{BVE3}, in this case only the
singlet component and the triplet one with the zero spin
projection on the direction of $\mathbf{M}_{e}$ exist in the
system. Both components penetrate into the ferromagnet over the
short distance $\xi _{F}$. If the $S/F$ interface transparency is
low {or the conductivity of the S film is much higher than the
conductivity of the F film}, the suppression of the order
parameter $\Delta $ is not essential and the superconducting
properties
remain almost unchanged.

The quantity $\delta N_{M}$ can be expressed in terms of the
quasiclassical normal Green function $\hat{g}$
\begin{equation}
\delta N_{M}=\sum_{p}(<c_{p\uparrow }^{\dagger }c_{p\uparrow
}-c_{p\downarrow }^{\dagger }c_{p\downarrow }>)=-i\pi \nu
T\sum_{\omega =-\infty }^{\omega =+\infty
}Tr(\hat{\sigma}_{3}\hat{g})  \label{M}
\end{equation}
where $\nu =p_{F}m/(2\pi ^{2})$ is the density of states at the
Fermi level, $\hat{\sigma}_{3}$ is the {third} Pauli matrix and
$\omega =\pi T(2n+1)$ is the Matsubara frequency. The normal Green
function $\hat{g}$ is a matrix in the spin space. In the
considered case of an uniform
magnetization it has the form $\hat{g}=g_{0}\cdot \hat{\sigma}%
_{0}+g_{3}\cdot \hat{\sigma}_{3}$. This matrix is related to the
Gor'kov anomalous matrix Green function $\hat{f}$ via the
normalization condition
\begin{equation}
\hat{g}^{2}-\hat{f}^{2}=1\;.  \label{NormCond}
\end{equation}
The matrix $\hat{f}$ describes the superconducting condensate. In
order to visualize how our results are obtained, we consider first
the simplest case when the condensate function $\hat{f}$ is small
in F and is close to its bulk value in the superconductor. {\ We
analyze the dirty case when the Usadel equation can be applied.
This means that s-wave superconductors are considered. These are
described by the standard BCS Hamiltonian with account for the
exchange field (in the ferromagnet) acting on the spins of the
free electrons.} {\ }{We write the Hamiltonian in the form}

\begin{equation}
\hat{H}=\hat{H}_{o}+\hat{H}_{S}+\hat{H}_{F}  \label{Ham}
\end{equation}
{Here }$\hat{H}_{o}${\ \ is the one-particle Hamiltonian which
includes the impurity scattering term }$U_{imp}$
\begin{equation}
\hat{H}_{o}=\sum_{\{p,s\}}\left\{ a_{sp}^{+}\left[ \xi _{p}\delta
_{pp^{\prime }}+U_{imp}\delta _{ss^{\prime }}\right] a_{s^{\prime
}p^{\prime }}\right\}  \label{Ham_o}
\end{equation}
{where }$\xi _{p}=p^{2}/2m-\epsilon _{F}${\ is the kinetic energy
counted from the Fermi energy }$\epsilon _{F}.${\ The second term
in Eq.(\ref {Ham}) is the standard BCS Hamiltonian for the
superconductor written in the mean-field approximation}
\begin{equation}
\hat{H}_{S}=-\sum_{\{p,s\}}\left\{ \Delta a_{\overline{s}\overline{p}%
}^{+}a_{sp}^{+}+c.c.\right\}  \label{Ham_S}
\end{equation}

{\ }{where }$\overline{s}=-s${\ (}$s=\pm 1)${\ \ and\ }$%
\overline{p}=-p.${\ The last term in Eq.(\ref{Ham}) describes the
ferromagnetic interaction in F. We accept the simplest form of
this part of the Hamiltonian (the mean-field approximation)}

\begin{equation}
\ \hat{H}_{F}=-\sum_{\{p,s\}}J\left\{ a_{sp}^{+}\mathbf{n}\ast \mathbf{%
\sigma }_{ss^{\prime }}a_{s^{\prime }p^{\prime }}\right\}
\label{Ham_F}
\end{equation}

{\ }{where the exchange energy is assumed to be positive for all
energies (the ferromagnetic type of interaction), }$n${\ is the
unit
vector parallel to the magnetization of the ferromagnet. The magnetization }$%
M${\ of the ferromagnet is proportional to the exchange energy }$J.$%
{\ If the contribution of free electrons strongly dominates (an
itinerant ferromagnet), one has }$M\cong M_{e}${. }

{If the polarization of the conduction electrons is due to the
interaction with localized magnetic moments, the Hamiltonian }$\hat{H}_{F}$%
{\ may be written in the form} {\cite{GR,BF,BEL}}

\begin{equation}
\ \hat{H}_{F}=-J_{1}\sum_{\{p,s\}}\left\{ a_{sp}^{+}\mathbf{S}\ast \mathbf{%
\sigma }_{ss^{\prime }}a_{s^{\prime }p^{\prime }}\right\}
\label{Ham_F1}
\end{equation}

{\ }{where }$S=\sum_{a}S_{a}\delta (r-r_{a}),${, }$S_{a}$%
{\ is the spin of a particular ion. A constant }$J_{1}${\ is
related to }$J${\ via the equation: }$J=J_{1}n_{M}S_{0}${\ ,
where }$n_{M}${is the concentration of magnetic ions and }$S_{0}$%
{\ is a maximum value of }$S_{a}${\ (we consider these spins as
classical vectors; see Ref. \cite{GR}). In this case the
magnetization is
a sum: }$\mathbf{M=M}_{loc}\mathbf{+M}_{e}${, and the magnetization }$%
\mathbf{M}_{e}${\ may be aligned parallel (}$J_{1}>0${, the
ferromagnetic type of the exchange field) \ to }$\mathbf{M}${\ or
antiparallel (}$J_{1}<0${, the antiferromagnetic type of the
exchange field). In the following we will assume a ferromagnetic
exchange interaction (}$\mathbf{M}_{e}$ {and }$\mathbf{M}${\ are
oriented in the same direction}$)$. {The case of antiferromagnetic
coupling will be briefly discussed below}. {In principle \ one can
add to Eq. (\ref
{Ham_F1}) the term }$\sum_{\{a,b\}}\left\{ \mathbf{S}_{a}\ast \mathbf{S}%
_{b}\right\} ${\ which describes a direct interaction between
localized magnetic moments {\cite{BF,BEL}}, but this term does not
affect final results. }

{Starting from the Hamiltonian (\ref{Ham}) and using a standard
approach {\cite{LO}}, one can derive the Usadel equation. In the
case of a low S/F interface transparency this equation can be
linearized. }Then, the function $\hat{f}$ {is} obtained from the
linearized Usadel equation (see e.g. Ref. \cite{BVE3})
\begin{equation}
\partial _{xx}^{2}f_{\pm }-\kappa _{\pm }^{2}f_{\pm }=0,\text{ in the F layer%
}  \label{FUsadel}
\end{equation}
and
\begin{equation}
\partial _{xx}^{2}\delta \hat{f}_{S}-\kappa _{S}^{2}\delta \hat{f}_{S}=%
\mathcal{K}(x)\hat{\sigma}_{3},\text{ in the S layer.}
\label{SUsadel}
\end{equation}
Here $\kappa _{\pm }^{2}=2(|\omega |\mp iJ\mathrm{sgn}\omega )/D_{F}$, $%
\kappa _{S}^{2}=2\sqrt{\omega ^{2}+\Delta ^{2}}/D_{S}$ and $\delta \hat{f}%
_{S}$ is a deviation of the function $\hat{f}_{S}$ from its bulk
(BCS) value
$f_{BCS}$, i.e. $\delta \hat{f}_{S}=\hat{f}_{S}-\hat{f}_{BCS}$, $\hat{f}%
_{BCS}=f_{BCS}\cdot \hat{\sigma}_{3}$. The functions $f_{F\pm }$
are the
elements (1,1) and (2,2) of the matrix $\hat{f}_{F}$. {\ The function $%
\mathcal{K}(x)$ contains the correction $\delta \Delta (x)$ to the
order parameter $\Delta $. This term is not relevant in our
calculations since only the component of $\hat{f}$ proportional to
$\hat{\sigma}_{0}$ contributes to the magnetization (see below,
Eq.(\ref{g3S})).} Eqs. (\ref {FUsadel}-\ref{SUsadel}) should be
complemented by the boundary conditions that can be written for
small $\hat{f}_{S,F}$ as
\begin{eqnarray}
\partial _{x}\delta \hat{f}_{S} &=&(g_{BCS}^{2}\cdot \hat{f}%
_{F}-g_{BCS}f_{BCS}\mathrm{sgn}\omega \cdot
\hat{\sigma}_{3})/\gamma _{S}
\label{SBC} \\
\partial _{x}\hat{f}_{F} &=&-(1/\gamma _{F})\hat{f}_{S}\;,  \label{FBC}
\end{eqnarray}
where $\gamma _{S,F}=R_{b}\sigma _{S,F}$, $R_{b}$ is the S/F
interface resistance per unit area, $\sigma _{S,F}$ is the
conductivity of the S or F region, and {\
$\hat{g}_{BCS}=g_{BCS}\cdot \hat{\sigma}_{0}$}. T{he BCS functions
have the well known form (see for example \cite{LO}) }

\begin{equation}
g_{BCS}=\omega /\sqrt{\omega ^{2}+\Delta ^{2}},f_{BCS}=\Delta
/i\sqrt{\omega ^{2}+\Delta ^{2}}  \label{BCS}
\end{equation}

The matrix function of the superconducting condensate $\hat{f}$
can be represented in the form
\begin{equation}
\hat{f}=f_{3}\hat{\sigma}_{3}+f_{0}\hat{\sigma}_{0}  \label{e2}
\end{equation}
for both regions. The component $f_{3} $ describes the singlet
condensate, whereas $f_{0}$ stands for the triplet component with
the zero projection of
the total spin of the pair on the direction of the magnetic moment $\mathbf{M%
}$. This functions are related to $f_\pm$ through: $%
f_{0,3}(x)=(1/2)(f_{+}(x)\pm f_{-}(x))$ The other components of
the triplet condensate arise only if $\mathbf{M}$ in the
ferromagnet is inhomogeneous.

Solving Eqs.(\ref{FUsadel}-\ref{SUsadel}) with the boundary
conditions, Eqs. (\ref{SBC}) and (\ref{FBC}), we find easily
\begin{eqnarray}
f_{F\pm }(x) &=&b_{\pm }\exp (-\kappa _{\pm }x)  \label{fF} \\
f_{S0}(x) &=&-a_{0}\exp (\kappa _{S}x)\;.  \label{f0S}
\end{eqnarray}
Here $b_{\pm }=\pm f_{BCS}/(\gamma _{F}\kappa _{\pm })$ and $%
a_{0}=g_{BCS}^{2}f_{F0}(0)/(\gamma _{S}\kappa _{S})$.
{\ As follows from Eqs. (\ref{fF}) and (\ref{f0S}) the functions
$f_{F\pm }$
and $f_{S0}(x)$ are small provided that $R_{F}/R_{b}<<1$ and $%
(R_{F}/R_{b})(R_{S}/R_{b})<<1$, where $R_{F,S}=\xi _{F,S}/\sigma
_{F,S}$ are the resistances (per unit area) of the F(S) of lengths
$\xi _{F,S}$.} In
order to calculate the magnetization we have to find the function $g_{3}=Tr%
\hat{\sigma}_{3}\hat{g}/2$ (see Eq. (\ref{M})). The latter is
related to the functions $f_{0,3}$ in the $F$ and $S$ region
through the normalization condition, Eq.(\ref{NormCond}), and is
given by
\begin{equation}
g_{F3}=f_{F0}f_{F3}\mathrm{sgn}\omega ,\;\;\;g_{S3}=f_{BCS}\delta
f_{S0}/g_{BSC}\;.  \label{g3S}
\end{equation}
As it has been discussed in Ref.\cite{BVE3}, the functions $f_{S0}$ and $%
f_{F0}$ corresponding to the triplet component of the condensate
are odd function of $\omega $ while the singlet components
$f_{BCS}$ and $f_{F3}$ are even functions. Thus, according to
Eqs.(\ref{g3S}) the functions $g_{F3}$
and $g_{S3}$ are even functions of $\omega $ ( $g_{BSC}$ is odd in $\omega $%
). This means that the sum over the frequencies in Eq.(\ref{M}) is
not zero and the proximity effect leads to a change $\delta M_{e}$
of the magnetization in both $F$ and $S$ layers (above $T_{c}$ the
magnetization in $S$ is zero).

After the qualitative discussion we have come to the conclusion
that the net magnetization due to the inverse proximity effect
must be negative. The explicit calculation based on Eqs.(\ref{M})
and (\ref{fF}) - (\ref{g3S}) confirms this result which is shown
in Fig. \ref{Fig.1} for some values of the parameters. We see that
$\delta M_{e}$ is negative, i.e the magnetization of the
ferromagnet is reduced and the superconductor acquires a finite
magnetization in the opposite direction. The change of the
magnetization $\delta M(x)$ extends over the length $\kappa
_{s}^{-1}$, which may be much larger than the thickness of the $F$
layer. This effect is another manifestation of the existence of
the triplet component of $\hat{f}$.

The inverse proximity effect considered here may be relevant for
several experiments on measurements of the magnetization in the
$S/F$ structures \cite{muhge,Garif}. In these experiments it was
found that the magnetization
started to decrease when crossing the superconducting critical temperature $%
T_{c}$ from above. The authors of these experiments compared the
data with the theoretical results of Refs. \cite{BEL,BB} that were
obtained under the assumption that the ferromagnetic order in the
$F$ thin layer might be modified due to the proximity effect
leading to the so called cryptoferromagnetic state. In these works
only the contribution of localized moments to the magnetization
was taken into account. Our calculations show that the conduction
electrons can give an additional contribution. The inverse
proximity effect leads to an additional reduction of the
magnetization $M_{e}$ and may serve as an alternative explanation
for the reduction of the magnetization observed experimentally
\cite{muhge,Garif}.

\begin{figure}
\includegraphics[scale=0.4]{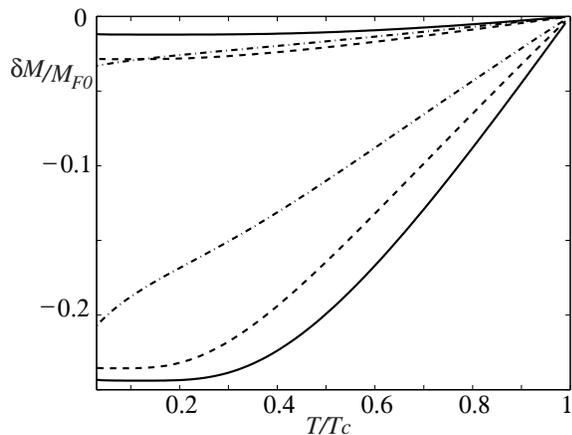}
\caption{Temperature
dependence of  $\delta M_S/M_{F0}$ (lower curves) and  $\delta
M_F/M_{F0}$ (upper curves) for the following values of
$\bar{\gamma}_F=\gamma _{F}/\xi_0$ ($\xi_0=\sqrt{D_{S}/2T_{c}}$):
$\bar{\gamma}_F=0.1$ (solid line), $\bar{\gamma}_{F}=0.3$ (dashed
line), and $\bar{\gamma} _{F}=0.5$ (dot-dashed line).  Here
$\gamma_F/\gamma_S =0.5$, $J/T_{c}=20$, $d_{F}/\protect\xi
_{0}=0.1$ and $\lambda=3$ (see [18])} \label{Fig.2}
\end{figure}

Let us analyze now an interesting case that may be relevant to the
experimental situation of Ref. \cite{Garif}. We assume that the
thickness of the $F$ layer $d_{F}$ is small compared to $\xi _{F}$
and that the Green
functions $g_{S}$ and $f_{S}$ are close to the bulk values $g_{BSC}$ and $%
f_{BCS}$. The latter assumptions is valid if the coefficient {\
$\gamma _{F}/\gamma _{S}=\sigma _{F}/\sigma _{S}$} is small
enough. In this case all functions in the $F$ region are not
necessarily small but they are almost
constant in space. Therefore we can average the exact Usadel equation over $%
x $ taking into account exact boundary conditions. Proceeding in
this way,
we get for the diagonal elements $g_{\pm }$ and $f_{\pm }$ of the matrices $%
\hat{g}$ and $\hat{f}$
\begin{equation}
g_{F\pm }=\tilde{\omega}_{\pm }/\zeta _{\omega \pm },\text{ \
}f_{F\pm }=\pm \epsilon _{bF}f_{BCS}/\zeta _{\omega \pm }
\label{gfPM}
\end{equation}
where $\tilde{\omega}_{\pm }\!\!=\!\!\omega +\epsilon _{bF}g_{BCS}\mp iJ$, $%
\zeta _{\omega \pm }=\sqrt{\tilde{\omega}_{\pm }^{2}-(\epsilon
_{bF}f_{BCS})^{2}}$, $\epsilon _{bF}=D_{F}/(2\gamma _{F}d_{F})$.
One can see that in the limiting cases of small and large energy
$\epsilon _{bF}$ the functions $g_{F\pm }$, $f_{F\pm }$ describe a
superconducting state with the energy gap equal to $\epsilon
_{bF}$ if $\epsilon _{bF}<<\Delta $ ( a subgap in the excitation
spectrum) and to $\Delta $ in the opposite case. In both cases the
position of the energy gap is shifted with respect to $\epsilon
=0$ (the Matsubara frequencies are related to $\epsilon $ via
$\omega =-i\epsilon $). It can be easily shown that the function
$g_{F3}$ that
determines the magnetization, Eqs.(\ref{M},\ref{g3S}), equals zero for $%
\epsilon _{bF}=0$ (very small S/F interface transparency) and for
very large
values of $\epsilon _{bF}$ (a perfect S/F contact). This dependence of $%
\delta M_{F}$ on $R_{b}$ leads to a nonmonotonic behavior of the
change of the magnetization {\ $\delta M_{F,S}$}. In
Fig.\ref{Fig.2} we show the temperature dependence of $\delta
M_{F,S}(0)$ for values of the parameters similar to those of
Ref.\cite{Garif}. We see that the decrease of the magnetization
may be of the order of 10\% and larger. This results correlate
with the experimental data of Ref.\cite{Garif} (see footnote
\cite{foot}). We have checked that $|f_{S0}|<1$ for the parameters
in Fig.{\ref{Fig.2}}.

{
We also present here analytical formulae for the ratio {\
$r_{S,F}=\delta M_{S,F}(0)/M_{F0}$ using Eq. (\ref{gfPM}) and
considering the case of low temperatures ($T<<\Delta $); $\delta
M_{S,F}$ are the magnetization variations in the $S$ and $F$ films
and $M_{F0}$ }is the magnetization in the $F$ film above $T_{c}$.
{The relation between }$M_{F0}${\ and }$J${\ depends on a
particular model of the ferromagnet. For example, in the simplest
model of the ferromagnet with a constant and
positive }$J${\ we have for an itinerant ferromagnet }{{$M_{F0}$}$%
=g\mu _{B}\nu J$ }(see \cite{foot}). To simplify the expressions for $%
r_{S,F} $ we assume also that $J<<\epsilon _{bF}\approx
(D_{F}/d_{F}^{2})(R_{F}/R_{b})(d_{F}/\xi _{F})$ {\ (this limit may
correspond to the experiment \cite{Garif})}. In this case we
obtain
\begin{eqnarray}
r_{S} &\approx &-1.67\sqrt{\Delta d_{F}^{2}/D_{S}} \\
r_{F} &\approx &-\pi \Delta /2\epsilon _{bF}  \label{r}
\end{eqnarray}
For estimations of the parameters one can take {\ experimental
values from Ref.\cite{Garif} where a ''weak'' ferromagnet
$Pd_{(1-x)}Fe_{x}$ was used. One gets $D_{F}/d_{F}^{2}=1000K$ for
$d_{F}=20A$. The Curie temperature which may be of the order of
$J$ varied from 90 to 250 K. The barrier (interface) resistance
$R_{b}$ is not known, but one can give a crude
estimation noting that $(R_{F}/R_{b})(d_{F}/\xi _{F})\approx T_{tr}$, where $%
T_{tr}$ is the transmission coefficient which varies from very
small values to a value of the order 1.} }

In conclusion, we have demonstrated the existence of the inverse
proximity effect in $S/F$ structures. Due to the presence of the
superconductor the magnetization in the ferromagnet {with the
ferromagnetic type of the exchange interaction} is reduced and a
magnetic moment is induced in the superconductor below $T_{c}$.
Its direction is opposite to the direction of the magnetic moment
in the ferromagnet and spreads over the superconducting coherence
length $\xi _{S}$. This distance can be much larger than the $F$
film thickness. The effect discussed may be the reason for a
reduced magnetization observed in $S/F$ structures leading to a
frequency shift of the magnetic resonance\cite{Garif}. {This
conclusion is changed in the case of ferromagnets with the
antiferromagnetic interaction between free electrons and localized
moments (negative }$J_{1}${\ in Eq.(\ref {Ham_F1})). If a
contribution of localized moments to the total magnetization in
such ferromagnets dominates (}$M_{loc}>M_{e}${), the
magnetisation }$\mathbf{M}_{e}${\ is opposite to }$\mathbf{M}${%
\ and therefore the induced magnetization variation in the superconductor }$%
\mathbf{\delta M}_{S}${\ will be parallel to }$\mathbf{M}.$

Note: After completing this work we became aware of the paper
Ref.\cite {valls2} where the magnetization leakage into the S
layer was numerically calculated for a ballistic S/F structure. In
this case the magnetization penetrates the S layer over distances
of the order of the Fermi wave length. We are not interested in
small scales of this order.

We would like to thank SFB 491 for financial support.


\begin{thebibliography}{20}
\bibitem{foot}  Note that we normalized $\delta M$ with respect to
the magnetization $M$ of the ferromagnet above $T_{c}$ assuming
that $M$ is determined by the free electrons (itinerant
ferromagnets). { In the case of ferromagnets with localized
magnetic moments and the ferromagnetic type
of interaction} all functions $\delta M/M$ should be multiplied by a factor $%
(1+\lambda )^{-1}$, where $\lambda =M_{loc}/M$ and $M_{loc}$ is
the contribution to the magnetization due to localized moments.
\end{thebibliography}
\end{document}